\documentstyle[prl,twocolumn,aps,epsf]{revtex}
\begin{document}
\draft

\twocolumn[\hsize\textwidth\columnwidth\hsize\csname
@twocolumnfalse\endcsname

\title{\bf   3D-2D  crossover   in  the   naturally  layered
superconductor (LaSe)$_{1.14}$(NbSe$_2$) }
\author{P. Samuely,$^1$ P.  Szab\'o,$^1$ J. Ka\v cmar\v c\'{\i}k,$^1$
A.   G.   M.   Jansen,$^2$   A.   Lafond,  $^3$  and  A.
Meerschaut$^3$, A. Briggs$^4$}
\address{$^1$Institute of Experimental Physics, Slovak Academy of
Sciences,         SK-04353         Ko\v{s}ice,         Slovakia.\\
$^2$Grenoble       High       Magnetic      Field      Laboratory,
Max-Planck-Institut     f\"{u}r     Festk\"{o}rperforschung    and
Centre  National  de  la  Recherche  Scientifique, B.P. 166,
F-38042 Grenoble Cedex 9, France.\\
$^3$  Institut des  Mat\'eriaux  Jean  Rouxel, BP  32229, F-44322
  Nantes Cedex 03, France.\\
$^4$ Centre de Recherche sur  les Tr\` {e}s
  Basses Temp\'{e}ratures,  CNRS, BP 166,  F-38042 Grenoble,
France }
\date{\today}

\maketitle

\begin{abstract}
 The temperature  and angular dependencies  of the resistive
upper critical magnetic field  $B_{c2}$ reveal a dimensional
crossover  of  the  superconducting   state  in  the  highly
anisotropic      misfit-layer     single      crystal     of
(LaSe)$_{1.14}$(NbSe$_2$)  with   the  critical  temperature
$T_c$  of 1.23  K. The  temperature dependence  of the upper
critical   field   $B_{c2\parallel   ab}(T)$   for  a  field
orientation  along  the  conducting  $(ab)$-planes  displays
a characteristic upturn at 1.1  K and below this temperature
the  angular dependence  of $B_{c2}$  has a  cusp around the
parallel field orientation. Both  these typical features are
observed  for  the  first  time  in  a naturally crystalline
layered system.
\end{abstract}

]

The  discovery of  the cuprate  superconductors with  a high
critical  temperature  has  lead   to  renewed  interest  in
anisotropic   or  even   low-dimensional  superconductivity.
Already  30   years  ago  Lawrence   and  Doniach  \cite{LD}
described  the   behavior  of  layered   superconductors  in
a magnetic  field  based  on   a  model  system  of  stacked
two-dimensional  superconducting layers  coupled together by
Josephson tunneling between adjacent  planes. In contrast to
the  isotropic case  where  the magnetic flux penetration occurs
in the form of vortices of circular
symmetry,  in  an anisotropic  superconductor   the
vortex  cores will be flattened in  the {\it  interlayer} $c$-direction
for magnetic fields parallel  to the layered  structure ($ab$-planes)
with  $\xi_{c}<\xi_{ab}$ for the vortex core radii $\xi_{c}$ and
$\xi_{ab}$, respectively, perpendicular and parallel to the
planes.
If $\xi_c$ is bigger than the
distance  between  the  adjacent  superconducting layers the
system is anisotropic but still three-dimensional with the
upper critical  magnetic fields  for fields perpendicular  and
parallel  to  the  layers   determined  by a product of  the corresponding
coherence  lengths:  $B_{c2||c}=\Phi_0/(2\pi\xi_{ab}^2)$ and
$B_{c2||ab}=\Phi_0/(2\pi\xi_{ab}\xi_c)$, respectively, where
$\Phi_0$  is  the  flux   quantum.
In some  cases  including  the  high-$T_c$  cuprates
the critical fields  can  be  very  high and
in  extremely anisotropic  superconductors   $\xi_c$
could  shrink with  decreasing  temperature
below  the value of the
interlayer  distance $s$. Then, the  vortices are  confined
between the  superconducting layers for fields applied
parallel   to  the   layers leading to a dimensional cross over
in the flux penetration.
Here, we present such a 3D-2D dimensional cross over in the upper
critical field data of the highly layered
single crystal of (LaSe)$_{1.44}$(NbSe$_2$). As shown in our
previous  paper  \cite{szabo}  (LaSe)$_{1.44}$(NbSe$_2$) has
been the first found system based on
conventional  superconductor behaving  like the  intrinsic
Josephson junctions.

According   to  the   simplest
Lawrence-Doniach model the upper critical $B_{c2||ab}$
parallel to the $ab$-planes is predicted to diverge
at the dimensional transition
at a temperature $T^*$, where  $\xi_c(T^*) = s/\sqrt{2}$.
The  real   finite  value  of   the  upper  critical   field
$B_{c2||ab}$ is  caused by the  finite superconducting layer
thickness,  Pauli  paramagnetism   and  spin-orbit  coupling
effects. Klemm, Luther and  Beasley \cite{klb} have extended
the  Lawrence-Doniach model  to include  these effects. They
show that the divergence is  removed but the dimensional crossover
to a  two-dimensional superconductivity is still characterized by
a strong upward  curvature of $B_{c2||ab}(T)$. Experiments
on the intercalated layered  compounds based on 2$H$-TaS$_2$
\cite{coleman}  revealed  strong  upward  curvature  of  the
temperature dependence of the parallel upper critical fields
accompanied  by  the  temperature  dependent  critical field
anisotropy $B_{c2||ab}/B_{c2||c}$ reaching maximum values of
about 60. Similar results  were obtained on the intercalated
MoS$_2$   \cite{woollam}.   In   experiments   performed  on
artificially grown superconducting superlattices
\cite{schuller,ruggiero}  the 3D-2D  crossover is  also well
expressed by a  kink  in  the  temperature  dependence  of
$B_{c2||ab}(T)$.

The  crossover  to  the   two-dimensional  behavior  can  be
highlighted  by a second  effect.
Namely, there  will be
a qualitative change of the  angular dependence of the upper
critical  field  $B_{c2}(\Theta)$.
  For the three-dimensional
situation,
the  angular  dependence $B_{c2}(\Theta)$ of the upper critical field
interpolates between  $B_{c2||ab}$ and $B_{c2||c}$
in the simple ellipsoidal form
$(\frac{B_{c2}(\Theta)\sin{\Theta}}{B_{c2||c}})^2                 +
(\frac{B_{c2}(\Theta)\cos{\Theta}}{B_{c2||ab}})^2   =  1   $  with
a rounded maximum  around $\Theta = 0$ (magnetic field parallel
to the layers).
 In  the  2D  regime  the
superconducting layers  are decoupled and can  be treated as
isolated  thin   films.  For this case Tinkham  \cite{tinkham}
proposed                     the                    equation
$|\frac{B_{c2}(\Theta)\sin{\Theta)}}{B_{c2||c}}|                  +
(\frac{B_{c2}(\Theta)\cos{\Theta}}{B_{c2||ab}})^2 = 1 $.
In the 3D  case the  slope of $B_{c2}(\Theta)$ is  zero around $\Theta = 0$
while  in the  2D case  it has
a finite value making a cusp.  This effect was also observed
in superconducting superlattices below the 3D-2D transition.
The situation  remains  unclear  in  the  natural layered single
crystals.  In the  case of  high-$T_c$ cuprates  the angular
dependence  of  the irreversibility  field   has  shown  up  the
cusp-like   form  for   the  most   anisotropic  system,
Bi$_2$Sr$_1$CaCu$_2$O$_x$  \cite{marcon,naughton}.  Here, we
present in the critical field data
of the naturally layered system (LaSe)$_{1.14}$(NbSe$_2$)
experimental  evidence  of  both  effects
pointing  to  the  3D-2D  transition:
The upturn  of the  upper critical field
parallel  to  the  layers  at  about  1.1  K  as well as the
cusp-like  behavior of  the angular  dependence of  $B_{c2}$
below this temperature.

(LaSe)$_{1.14}$(NbSe$_2$)     is    a     low    temperature
superconductor  with $T_c$  around  1.2  K belonging  to the
family  of  the  lamellar  chalcogenides \cite{meerschaut1},
where  the two slabs, LaSe and NbSe$_2$, are  stacked in  a certain
sequence.  Due  to  the  different  symmetry  of  the LaSe and
NbSe$_2$  layers   a  crystallographic misfit  results
along  one  intralayer axis whereas along  the  perpendicular
intralayer axis a perfect fit of both structures is achieved
\cite{meerschaut1}.        In       the        case       of
(LaSe)$_{1.14}$(NbSe$_2$) every intercalated LaSe layer with
the  thickness  of  about  6~\AA ~is  sandwiched  by  one
2H-NbSe$_2$  layer  with  about   the  same  thickness.  The
sandwich unit  is stabilized by  the electron transfer  from
the  LaSe  to  the  NbSe$_2$  slab  resulting in the natural
layered system of insulating LaSe and (super)conducting
NbSe$_2$ sheets, where the conduction is accomplished by the
Nb  4d$_{z^2}$  orbitals  \cite{meerschaut1,berner}.

A standard lock-in  technique at 17  Hz was used  to measure
the magnetic field dependencies  of the sample resistance in
a four-probe  configuration at  different fixed temperatures
down to 100  mK. The current and voltage  was measured within
the  same  $ab$-plane  of  the  sample. The magnetoresistive
transitions  were measured  at different  angles between the
$ab$-plane  of the  sample  and  the applied  magnetic field
keeping  the current  always  orthogonal  to the  field. The
angular resolution  was better than 0.2  degree with the $\Theta
= 0^{\circ}$ orientation  defined from the  highest value of
$B_{c2}$.

Figure  1 shows  the magnetoresistive  transitions from  the
superconducting  to normal  state for  the applied  magnetic
field  oriented  perpendicular  (upper  part)  and  parallel
(lower  part) to  the superconducting  layers. The  measured
magnetoresistances were normalized to  the high field value.
The transitions  are shifted to higher  fields and broadened
with decreasing temperature. After the main increase of
the  resistance  there  is  a  large  field  range where the
resistance   is  slightly   increasing  before   it  reaches
the constant normal-state value. This  effect is  more remarkable  for the
perpendicular field orientation indicating the important role
of  fluctuations  of  the  phase  and  the  amplitude of the
superconducting  order  parameter  in  the  system.  In such
a situation the magnetoresistive transitions are affected by
dissipation effects due to
depinned  vortices  much  before  the  mean field transition
$B_{c2}$  is  achieved.  To  avoid  this  effect  as much as
possible we have defined the  critical fields near the onset
of  superconductivity where  the resistance  is $R/R_N$=0.95
($R_N$ - normal state resistance).

Figure 2 shows the  resulting upper critical magnetic fields
for   both    field   orientations   $B_{c2||c}$   and
$B_{c2||ab}$,
respectively. The parallel-to-the-plane upper  critical magnetic field
shows a  strong upward curvature
below 1.1 K.  At 100 mK it reaches a  value of 18 Tesla, what
is far above the  Clogston paramagnetic limit. $B_{c2||c}$
perpendicular  to  the  layers  reveals  a slightly positive
curvature  below $T_c$  down to  400 mK  and a saturation at
the lowest  temperatures. A  positive curvature  contradicts the
expected linear behavior in this temperature range predicted
for  classical  type-II  superconductors  \cite{whh}. On the
other  hand  this  feature  can  be  found  in many "exotic"
superconductors   including   the   layered  dichalcogenides
\cite{brandow}.  Among  various  explanations  we  remark an
effect  of  the  flux  flow  when  the  vortices melt due to
important phase fluctuations in systems with a high value of
the upper critical field or $dB_{c2||c}/dT$ at $T_c$ \cite{tesanovic}.
In our case
it is more than 1 Tesla/Kelvin \cite{szabo}.
The plausibility  of such an explanation is also
supported  by  the  fact  that  the  effect  of the positive
curvature    is    much    stronger  for another misfit-layer
system, (LaSe)$_{1.14}$(NbSe$_2)_2$ ($T_c$=5.3 K),  with smaller
superconducting coherence volume and consequently  stronger fluctuations
\cite{kacmarcik1}.

We  note  that  the  real   $B_{c2}$  values  can  be
significantly higher than those defined at 0.95 $R_N$. In
\cite{kacmarcik2} we applied the model of Ullah and Dorsey \cite{ullah}
for  a scaling analysis
 of the contribution of superconducting fluctuations
 to the magnetoresistive transitions
 of (LaSe)$_{1.14}$(NbSe$_2)$
 in perpendicular fields. Using the upper critical field
 as the only fitting parameter in the
 scaling analysis we have obtained  a classical temperature dependence of
$B_{c2||c}(T)$ with  a linear decrease towards to $T_c$  and a
zero-temperature extrapolated $B_{c2}(0)$ equal to 1.2 Tesla
instead  of 0.4 T  obtained at 0.95  $R_n$.
An additional argument for  the fact that the field value of
$B_{c2}$  lies  far  behind  the  main  increase of the {\it
intralayer}  resistance has been
found  in our  previous  paper  \cite{szabo}. There  we have
shown    that    the    {\it    interlayer}   transport   in
(LaSe)$_{1.14}$(NbSe$_2)$  is accomplished  by the tunneling
of  quasiparticles  and  Cooper  pairs  across the Josephson
coupled layers.  At higher magnetic fields the tunneling of
quasiparticles  dominates  with   a  linear  magnetic  field
dependence   of  the   interlayer  conductance   until  this
conductance achieves the constant normal-state value at a certain magnetic
field defining the value  of
$B_{c2||c}(T)$.  The resulting  temperature  dependence  of  the
upper critical  magnetic field is  in a full  agreement with
the Werthamer  prediction and gives  $B_{c2||c}(0)$ equal to
1.2 Tesla in agreement with the critical fluctuations model.
Therefore,  the estimation of the upper critical  field
from the {\it intralayer} resistance  measurements in the present
experiment  inevitably  interferes  with  the
dissipative vortex  processes and can  be taken only  as a
lower bound of the  real upper critical fields.

The  upper critical  field anisotropy $\gamma = B_{c2||ab}/B_{c2||c}$
as a function of the temperature  is shown in the lower part
of Fig. 2. Just below  $T_c$ it starts at a value  of $\sim $ 40
and rises rapidly to a maximum  of $\sim$ 130 at around 1000
mK  and  it  smoothly  decreases  down  with  a  tendency to
saturation  at  the  lowest  temperatures  where it achieves
a value    of    $\sim$    50.    The    rapid    rise    of
$B_{c2||ab}/B_{c2||c}$ below 1.2 K demonstrates the dominant
effect of the upturn of  $B_{c2||ab}$ compared to the
anomalous   positive  curvature   of  $B_{c2||c}(T)$.   Even
assuming $B_{c2||c}(T)$   to be linear
between $T_c$ and 400 mK  would not have an important
 effect on the
overall  temperature dependence  of the  anisotropy factor
with a maximum at about 1.0~K, just the $\gamma$ values would be
lowered in this temperature range.
This  supports a dimensional crossover around 1~K.

Just   below   $T_c$   the   three-dimensional   anisotropic
Ginzburg-Landau limit is valid in any superconductor because
the divergent temperature dependence of the coherence length close to $T_c$
will finally increase $\xi_c$
above the interlayer distance $s$ ($\xi_c > s$). In our interlayer
transport measurements  we found \cite{szabo}  that the real
$B_{c2||c}(1.2~{\rm K})$  is  about   0.2~T giving an
in-plane coherence length $\xi_{ab}  \sim 400$~\AA. Taking
into account  the anisotropy factor  at this temperature  we
can  estimate  the  out-of-plane coherence  length  $\xi_c  =
\xi_{ab}/\gamma \approx 10$~\AA. This value is comparable with
the $c$-axis lattice constant ($\sim 12$~\AA) what indicates
a realistic  starting condition  for the  3D-2D dimensional
transition since at lower  temperatures $\xi_c$ becomes
smaller than $c$ and the flux can be trapped in between the
superconducting  layers  making  the  orbital  pair-breaking
effect ineffective.

Below the  3D-2D transition the  anisotropic Ginzburg-Landau
limit is no longer valid and the interlayer coherence length
$\xi_c$  can not  be  anymore  calculated from  the parallel
upper critical field  $B_{c2||ab}$ since the superconducting
layers are almost decoupled.  For the single superconducting
layer  of the  thickness  $d$,  the parallel  upper critical
field    is    given    \cite{degennes}    as   $B_{c2||ab}=
\Phi_0\sqrt{12}/(2\pi d\xi_{ab})$.  In  a  more  realistic  model
Deutscher and Entin-Wohlman \cite{deutscher} have shown that
a layered    superconductor   consisting    of   alternating
insulating  layers  of  thickness  $s$ and (super)conducting
layers of  thickness $d$ shows a  dimensional crossover with
an upturn  in $B_{c2||ab}(T)$ when  the interlayer coherence
distance $\xi_c$ is
smaller than  the sum of $s  + d$. They have  found that the
two-dimensional  behavior  of  a   layered  system  is  more
pronounced  when the  insulating layer  thickness is  bigger
than  the superconducting  layer  thickness.
Being  in the 2D regime,
the  temperature dependence of  the parallel critical
field  can be directly compared with that of the  square  root
of the  perpendicular critical  field.  This  is  demonstrated  by  the open
symbols   in   the   Fig. 1   giving $\sqrt{B_{c2||c}(T)}$
multiplied with a constant factor.
At  temperatures  above  1  K  one  can see
disagreement between the  temperature
dependence of  $B_{c2||ab}$ and of $\sqrt{B_{c2||c}(T)}$
indicating the  crossover to the
three  dimensional regime.  Of coarse  such a  comparison is
a significant oversimplification since in  our case also the
spin-orbit scattering plays an important role producing very
high upper critical fields in the $ab$-plane.

Another  strong   argument  for  the   3D-2D  transition  of
a superconductor  can be  obtained via  studying the angular
dependence of the upper critical field. In Fig. 3 we present
the  data  for four   different  temperatures at
0.45, 0.58, 1, and 1.2 K.
The data are compared with  both models: the anisotropic
3D model of Lawrence and Doniach  as well as the 2D model of
Tinkham.  Since for  such  high degree  of anisotropy  the
models differ  only for the field  orientations very near to
the $ab$-planes, the data are  presented only at small range
of  angles   $\Theta$.  As  is  apparent   the  behavior  is
cusp-like  for the  temperatures $T$   = 0.45  K and  0.58 K
indicating    the   two-dimensional    character   of    the
superconducting state. At
$T$ =  1 K these two  models are almost nondistinguishable
due to  the high anisotropy factor,  but still more probably
the system  is two-dimensional as no  round maximum could be
observed in $B_{c2}(\Theta)$. At $T$ = 1.2  K a cusp-like
form  near $\Theta$  = 0   is changed  to the  round maximum
indicating more three-dimensional behavior.
The values of $B_{c2||ab}$
diminish  as temperature increases in a similar way as the
data presented in Fig. 2 for a different crystal. Also the
anisotropy factor  behaves qualitatively as in  the
Fig. 2, i.e. it increases from the lowest temperature to the
3D-2D crossover and then it drops down in the 3D regime near
$T_c$. We
note that the value of the anisotropy factor  differs from
sample to  sample \cite{kacmarcik3}, but the  qualitative picture holds.  Such
a pronounced  cusp-like  behavior of $B_{c2}(\Theta)$   in  the  two-dimensional
regime could be observed only  on very tiny samples with
a clear plane parallel geometry,  while in samples with little
wavy  surfaces the  angular dependence  always revealed  the
round maximum  which could be  fitted by the  3D anisotropic
formula.

Finally,   in   the   strongly   anisotropic  superconductor
(LaSe)$_{1.14}$(NbSe$_2)$ with $T_c$ =  1.23 K we have found
experimental   evidence   for   the   crossover   from   the
three-dimensional    to    two-dimensional   superconducting
behavior.  Namely,  we  observed  a  strong  upturn  in  the
temperature  dependence  of   the  parallel  upper  critical
magnetic   field  at   1.1~K with, at lower temperatures,
a temperature  dependent critical field
anisotropy  and  a cusp-like angular dependence of
$B_{c2}$ around the parallel field
orientation. All these typical features are observed for the
first time in a naturally crystalline layered system.

Supports of the  Slovak VEGA  grant
No.1148 and the HPP
Programme  "Transnational   Access    to   Major   Research
Infrastructures"  are greatly acknowledged.

\newpage
\onecolumn

  \begin{figure}
\epsfverbosetrue
\epsfxsize=10cm
\epsfysize=16cm
   \hspace{5mm}
  \begin{center}
\epsffile{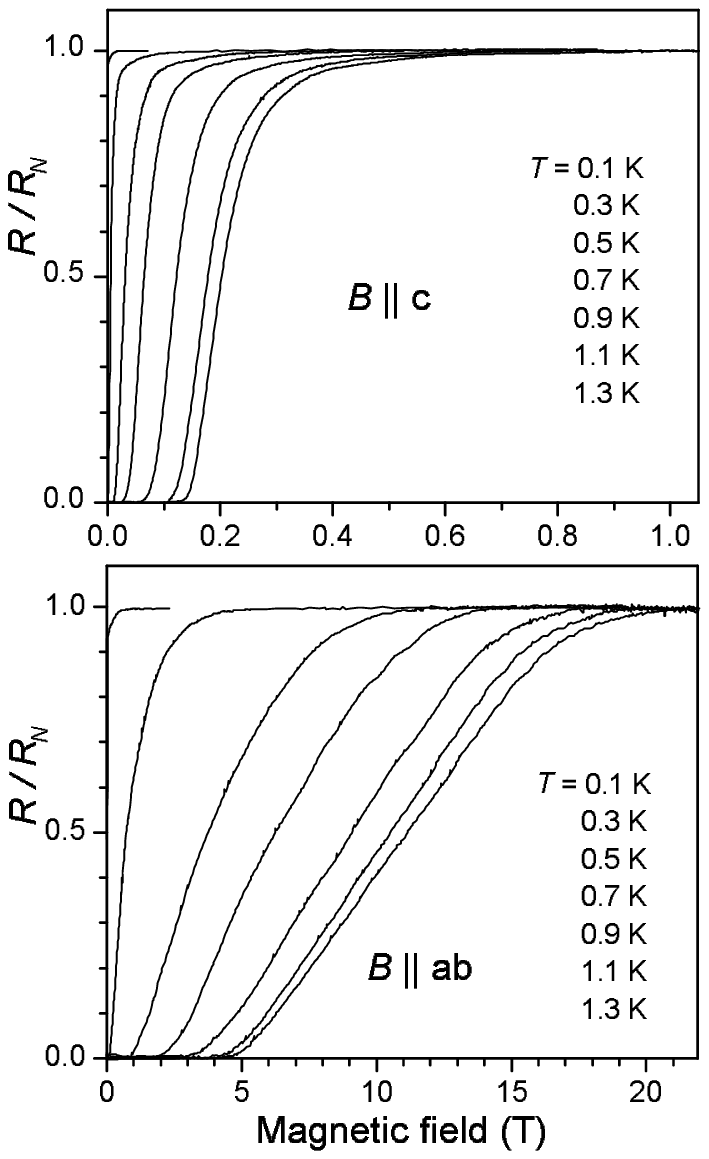}
 \end{center}
\vspace{30mm}
\caption{
Magnetoresistive
  superconducting transitions at  different temperatures for
the  applied  field  perpendicular   (upper part) and parallel to  the  superconducting
planes .
}
\end{figure}

\newpage

\begin{figure}
\epsfverbosetrue
\vspace{3mm}
\epsfxsize=10cm
\epsfysize=16cm
\hspace{3mm}
\epsffile{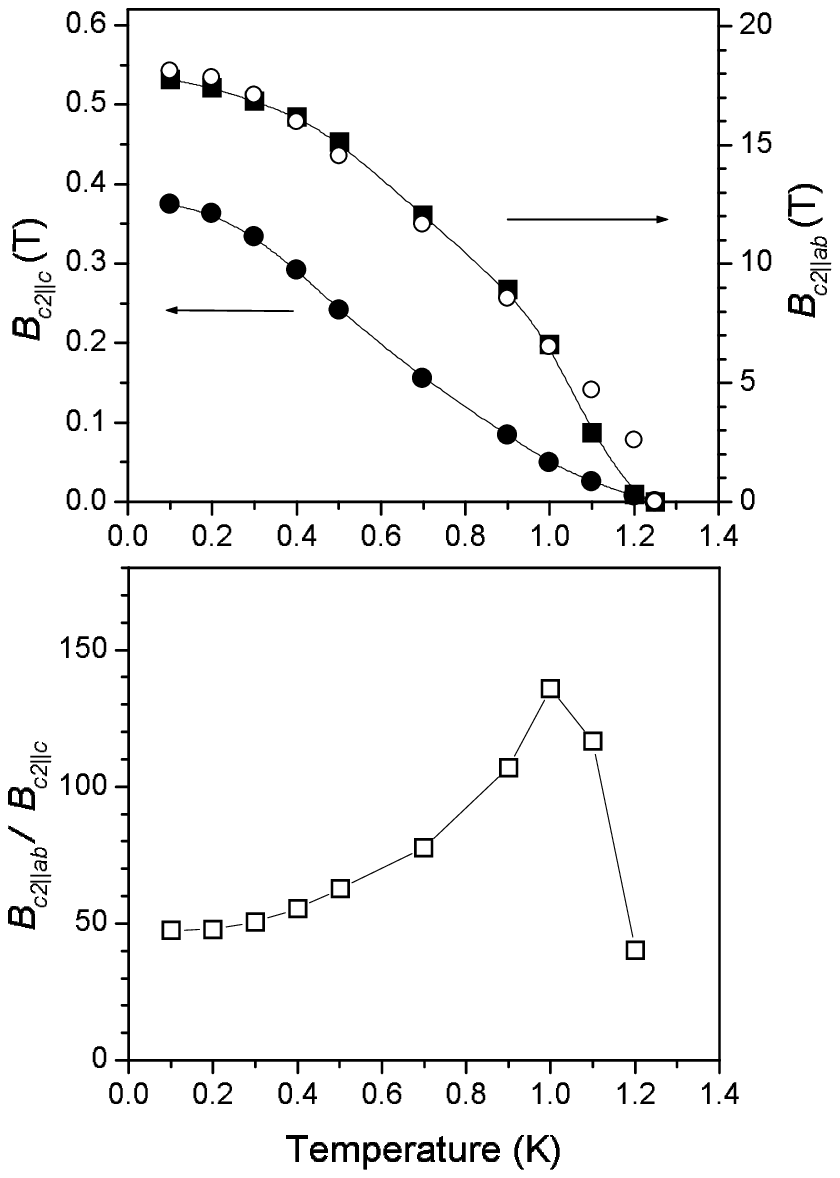}
\vspace{30mm}
\caption{
Upper part  - resistive upper  critical fields for  $B || c$
and $B  || ab$ are  shown by closed  symbols.
$B_{c2||ab}$  calculated  from  $B_{c2||c}$  in  2D  model -
opened symbols.  Lower part - the  temperature dependence of
the anisotropy factor.
}
\end{figure}

\newpage

\begin{figure}
\epsfverbosetrue
\vspace{3mm}
\epsfxsize=14cm
\epsfysize=13cm
\hspace{3mm}
\epsffile{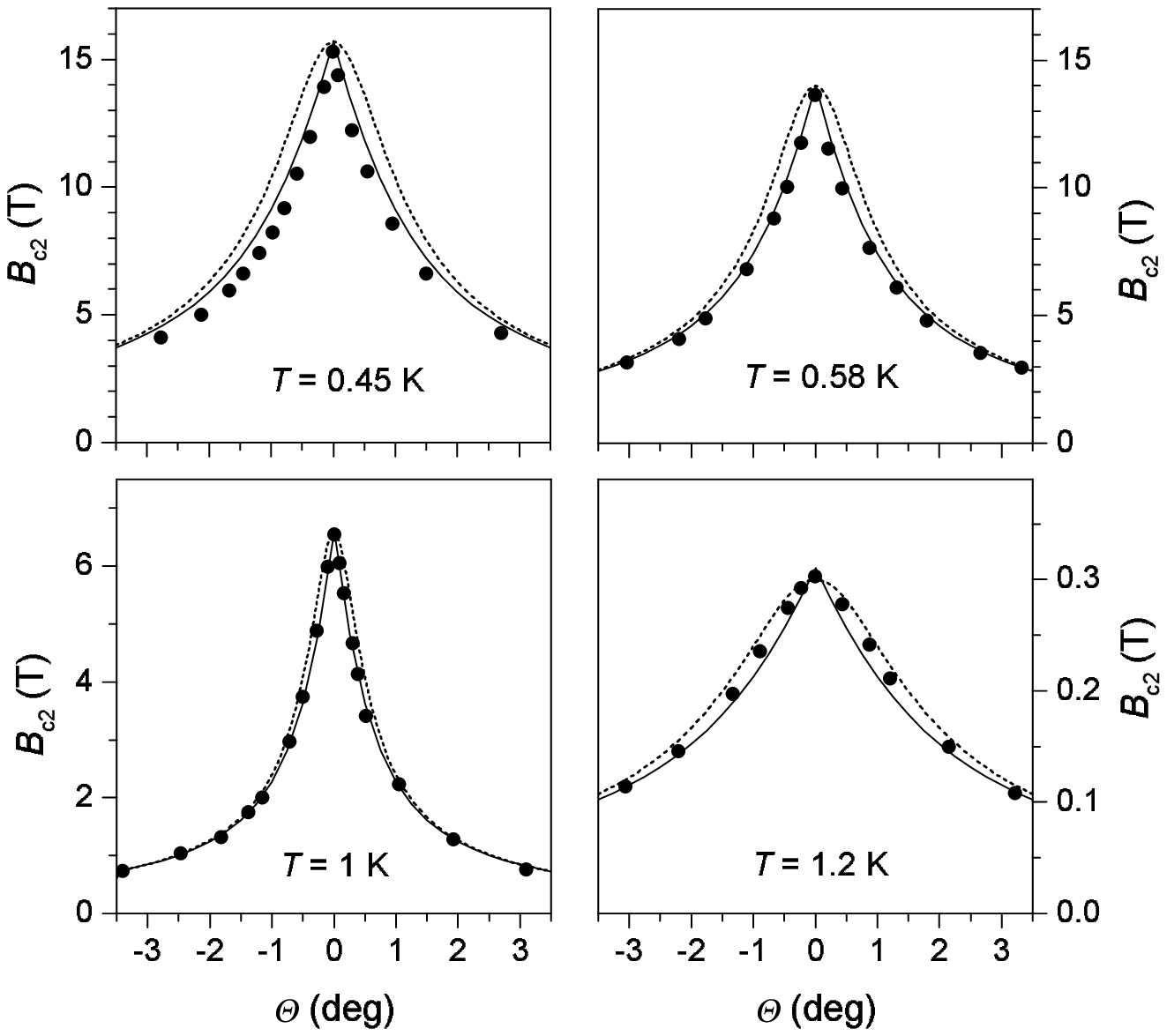}
\vspace{30mm}
\caption{
The  angular  dependence  of  the  upper  critical  field at
different temperatures.  Solid line - the  2D Tinkham model,
dashed line - the anisotropic 3D Lawrence-Doniach model.
}
\end{figure}

\end{document}